\documentclass[runningheads,a4paper]{llncs}
\usepackage{amssymb}
\usepackage{graphicx}
\usepackage{url}
\usepackage{multirow}
\urldef{\mailsa}\path|{lavrentyeva, novoselov, simonchik}@speechpro.com|    
\newcommand{\keywords}[1]{\par\addvspace\baselineskip
\noindent\keywordname\enspace\ignorespaces#1}

\usepackage[utf8]{inputenc}
\usepackage[T2A]{fontenc}

\begin{document}
\mainmatter  

\title{Anti-spoofing Methods for Automatic Speaker Verification System}

\titlerunning{Anti-spoofing Methods for Automatic Speaker Verification System}

%
%
\author{Galina Lavrentyeva\inst{1,2} \and Sergey Novoselov\inst{1,2} \and Konstantin Simonchik\inst{1,2}}
\authorrunning{Anti-spoofing Methods for Automatic Speaker Verification System}

\institute{Speech Technology Center Limited, St.Petersburg, Russia \\
\url{www.speechpro.com} \\
\and
ITMO University, St.Petersburg, Russia \\
\url{www.ifmo.ru} \\
\mailsa\\
}

%
%

\toctitle{Lecture Notes in Computer Science}
\tocauthor{Authors' Instructions}
\maketitle

\begin{abstract}
Growing interest in automatic speaker verification (ASV) systems has lead to significant quality improvement of spoofing attacks on them. Many research works confirm that despite the low equal error rate (EER) ASV systems are still vulnerable to spoofing attacks. In this work we overview different acoustic feature spaces and classifiers to determine reliable and robust countermeasures against spoofing attacks. We compared several spoofing detection systems, presented so far, on the development and evaluation datasets of the Automatic Speaker Verification Spoofing and Countermeasures (ASVspoof) Challenge 2015. Experimental results presented in this paper demonstrate that the use of magnitude and phase information combination provides a substantial input into the efficiency of the spoofing detection systems. Also wavelet-based features show impressive results in terms of equal error rate. In our overview we compare spoofing performance for systems based on different classifiers. Comparison results demonstrate that the linear SVM classifier outperforms the conventional GMM approach. However, many researchers inspired by the great success of deep neural networks (DNN) approaches in the automatic speech recognition, applied DNN in the spoofing detection task and obtained quite low EER for known and unknown type of spoofing attacks.
\end{abstract}

\section{Introduction}

Biometrics technologies play an essential role in restricting access to informational resources in today's world. One of the reliable approaches of guarding access to important data is speaker recognition. Speaker recognition systems are widely used in customer identification during call to a call center, passive identification of a possible criminal using a preset "black list", Internet-banking systems and other fields of e-commerce.\\
Automatic speaker verification systems aim to detect if the utterance belongs to the real speaker registered in the system or to the impostor.  Although performance of automatic speaker verification (ASV) techniques has improved in recent years, they are still acknowledged to be vulnerable to spoofing attacks.\\
There are two types of spoofing attacks on the ASV systems: direct attack and indirect attacks. Indirect attacks require access permission to the system and can be applied to the inner modules (feature extraction module, voice models or classification results), while direct attacks focus only on the input data and are more likely to be used by criminals due to implementation simplicity. The most well-known spoofing attacks are “Impersonation”, “Replay attack”, “Cut and paste” \cite{villalba}. But the most threatful are speech synthesis and voice conversion approaches. Voice conversion is the process of modifying a speech signal of the source speaker to sound like the target speaker. Speech synthesis is the computer-generated simulation of human speech.\\
Despite the development of new robust spoofing detection methods, most of them depend on a training dataset related to a specific spoofing attack. In real cases the nature of spoofing attack is unknown, that is why generalized spoofing detection methods are very important \cite{ASV}. That was the motivation for researches from University of Eastern Finland to organize the Automatic Speaker Verification Spoofing and Countermeasures (ASVspoof) Challenge 2015 \cite{ASV} in order to support the development of new spoofing detection algorithms, where we also presented our systems for spoofing detection and achieved 2nd result.\\
In this paper we concentrate on the investigating the most appropriate front-end features and classifiers for the spoofing detection system, which is effective in stand-alone spoofing detection task. In particular, we investigated anti-spoofing systems (ASS) introduced on the ASVspoof Challenge 2015 and compared results proposed by its authors. The aim of our research was to find the most effective method for detecting unknown spoofing attacks.

\section{ASVspoof Challenge 2015}
ASVspoof Challenge was organized by Zhizheng Wu, Tomi Kinnunen, Nicholas Evans and Junichi Yamagishi from University of Eastern Finland in 2015 to encourage the research work in spoofing detection field and stimulate the development of generalised countermeasures. According to \cite{ASV} the main aim of the Challenge was to generalize the proposed spoofing detection systems on the base of their vulnerability results on one common database with varying spoofing attacks. The data set includes genuine and spoofed speech generated by 10 different spoofing algorithms using voice conversion and speech synthesis. The main purpose was to provide an opportunity to develop generalized countermeasures trained on the known type of attacks and test them  on the unknown spoofing attacks.
\subsection{Training, Development and Evaluation Data sets}

The training data set contains 3750 genuine and 12625 spoofed utterances collected from 25 speakers(10 male, 25 female). The development data set consists of 3497 genuine and 49875 spoofed trials from 35 speakers (15 male, 20 female). To generate spoofed utterances 5 spoofing methods (called known attacks) \cite{ASV} were selected because of their simple implementation:
\begin{itemize}
    \item S1 - simplified frame selection algorithm, based on voice conversion. The converted speech is generated by selecting target speech frames
    \item S2 - voice conversion algorithm which adjusts the first mel-cepstral coefficient to shift the slope of the source spectrum to the target
    \item S3, S4 - speech synthesis system based on Hidden Markov model with speaker adaptation techniques by 20(S3) and 40(S4) adaptation utterances
    \item S5 - voice conversion (using voice conversion toolkit and Festvox system)
\end{itemize}
The evaluation data set contains 9404 genuine and 184000 spoofed utterances from 46 speakers(20 male and 26 female). Spoofed trials were generated by 5 methods, used for development and training sets and additional 5 spoofing methods for unknown attacks \cite{ASV}. The additional methods were: 
\begin{itemize}
    \item S6 - voice conversion algorithm based on joint density GMM and maximum likelihood parameter generation considering global variance
    \item S7 - voice conversion algorithm  similar to S6, using line spectrum pair for spectrum representation
    \item S8 - tensor-based approach to voice conversion, using Japanese set for speaker space construction
    \item S9 - voice conversion algorithm which uses kernel-based partial least square to implement a non-linear transformation function
    \item S10 - speech synthesis by open-source MARY TTS
\end{itemize}
\section{Front-end}
The main components of spoofing detection system are feature extraction and decision making modules. However some participants of the ASVspoof Challenge 2015 used additional steps in their systems, such as front-end preprocessing and high level features extraction.
\subsection{Front-end preprocessing}
There were several signal preprocessing techniques proposed in the spoofing detection systems. The first purpose of these techniques is to to enhance the impact of different features on the spoofing detection system decision, while the second is to detect simple types of spoofing attack by some enormous for natural speech artifacts and eliminate these utterances in further analysis. 
\subsubsection{Pre-detector.}
After experiments on the training part of the challenge database we decided to include pre-detection as a preliminary step in our spoof detection system \cite{B}. The pre-detector checks whether the input speech signal has zero temporal energy values. In case of zero-sequence the signal is declared to be a spoofing attack, otherwise the speech signal is used as input data for the feature extractor. But the significant limitation of the described pre-detector is that it will be useless in case of channel effects or additive noise.

\subsubsection{Bandpass filter.}
Various experiments were made for acoustic features extracted from signal of different frequencies. These experiments show that different features are more informative on different ranges of frequencies. That is why we decided to check bandpass filter that rejects frequencies outside the specific for the type of features range. The results of these experiments for one of our spoofing detection systems based on phase-based features, described in next section are presented in Table \ref{tab:bandpass}.
\begin{table}[]
    \centering
    \begin{tabular}{|c|c|c|c|c|c|c|}
        \hline
        \multirow{2}{*}{Frequency range (Hz)} & \multicolumn{6}{ c| }{Spoofing type}\\
        \cline{2-7}
        & S1 & S2 & S3 & S4 & S5 & All\\
        \hline	
        0 - 3400 & 2.71 & 4.28 & 0.23 & 0.26 & 1.87 & 2.36\\
        0 - 8000 & 2.13 & 4.3 & 0.77 & 0.74 & 3.39 & 2.51\\
        \hline	
    \end{tabular}
    \caption{Bandpass filter effect on EER of TV-SVM spoofing detection system with phase-based features (\%)}
    \label{tab:bandpass}
\end{table}
\vspace{-1cm}
\subsubsection{Pre-emphasis.}
Pre-emphasis refers to filtering that emphasizes the higher frequencies and downplay the lower ones. Its purpose is to balance the spectrum of voiced sounds that have a steep roll-off in the high  frequency region. Pre-emphasis removes some of the glottal effects from the vocal tract parameters. Comparison results of \cite{A} demonstrates that the state-of-the-art Mel frequency cepstral coefficients (MFCC) are sensitive to pre-emphasis. They are presented in Table \ref{tab:preemph} and illustrate the usefulness of the pre-emphasis step for anti-spoofing purposes.
\begin{table}[]
    \centering
    \begin{tabular}{|c|c|c|c|}
         \hline
         & MFCC & MFCC + $\Delta$ & MFCC + $\Delta$ + $\Delta\Delta$\\
         \hline
         No Pre-emphasis & 4.00 & 2.66 & 2.80\\
         Pre-emphasis $\alpha$ = 0.97 & 3.26 & 2.17 & 1.60\\
         \hline
    \end{tabular}
    \caption{Pre-emphasis performance on the base EER of GMM spoofing detection system on the development dataset (\%)}
    \label{tab:preemph}
\end{table}
\vspace{-1cm}
\subsubsection{Voice Activity Detection.}
In order to discard useless information from the speech signal several participants tried to use Voice Activity Detector (VAD) as the preprocessing step for their spoofing detection systems. Authors of \cite{C} apply DNN-based VAD and remove only first and last non-speech fragments. In \cite{D} authors offered to use pitch based VAD on the score extraction level to discard scores of all silence patches, each of which contains 51 feature frames (with 0.025s frame length and 0.01s frame shift) and covers about 0.5s of temporal context. In \cite{E} authors remove all non-speech fragments by GMM-based VAD \cite{VAD_E1}.\\
However experiments with our systems show that using VAD segmentation for full utterance is ineffective. According to our opinion, applying VAD may lead to throwing out informative artifacts locating between speech fragments. It is confirmed by the results of comparison for two TV-SVM systems with MFCC features performed in Table \ref{tab:VAD}.\\
\begin{table}[]
    \centering
    \begin{tabular}{|c|c|c|c|c|c|c|}
        \hline
        & \multicolumn{6}{ c| }{Spoofing type}\\
        \cline{2-7}
        Preprocessing type & S1 & S2 & S3 & S4 & S5 & All\\
        \hline	
        no VAD & 4.91 & 19.56 & 0.7 & 0.86 & 7.87 & 8.66 \\
        VAD & 8.51 & 30.06 & 4.86 & 5.03 & 8.04 & 13 \\
        \hline	
    \end{tabular}
    \caption{Effect of VAD on the spoofing detection performance on the base of EER for TV-SVM system with MFCC features on the development dataset (\%)}
    \label{tab:VAD}
\end{table}
\vspace{-1cm}
\subsubsection{Resampling.} Signal preprocessing in \cite{K} includes downsampling original signal recordings from 16 kHz to 8 kHz to reduce computational load. In this case computational time greatly reduces, but our experiments show that during the downsampling process essential information is loosing which affects the performance of spoofing detection.
\subsection{Front-end features}
Most of the participants of the ASVspoof Challenge 2015 found out the efficiency of the front-end features obtained by fusion of features appropriate for detecting specific spoofing attack. Thus, acoustic feature extractors in proposed systems are combinations of two or more different acoustic feature extraction methods. The most powerful features were attained by combining magnitude and phase information. It is hard to present full comparison of the implemented features because of the different type of classifiers used after, but we can analyse, how powerful are proposed approaches for spoofing detection task.
\subsubsection{Magnitude based features.}
The magnitude spectrum contained detailed information about speech signal. Previous works has demonstrated the usefulness of magnitude information for spoofing detection task \cite{handbook}. Most part of systems proposed during the ASVspoof Challenge used magnitude based features with and without their derivatives.\\
Most of the successful spoofing countermeasures use \textbf{\emph{Mel-frequency Cepstral Coefficients}} (MFCC) with their first and second derivatives as acoustic level features. They were used in \cite{B}, \cite{E}, \cite{G}, \cite{I} and \cite{J}.\\
In \cite{J} authors proposed \textbf{\emph{Linear-frequency Cepstral Coefficients}} (LFCC) by using linear filterbank instead of mel-filterbank.\\
We also use \textbf{\emph{Mel-frequency Principle Coefficients}} (MFPC) coefficients, that were obtained similar to MFCC coefficients, but using principal component analysis instead of the discrete cosine transform to achieve decorrelation of the acoustic features\cite{B}. Table \ref{tab:MWPC} presents EER of spoofing detection performance for the development dataset for MFCC and MFPC-based spoofing detection systems.
These results demonstrate that we achieved substantial EER improvement for all spoofing techniques by PCA basis implementation.\\
Another approach to use magnitude information is extraction of \textbf{\emph{Log Magnitude Spectrum}} features (LMS) and \textbf{\emph{Residual Log Magnitude Spectrum}} features (RLMS)\cite{D}. Table \ref{tab:MGDF} shows comparison results for systems using these features.

\subsubsection{Phase-based features.}
Most approaches to detect synthetic or voice converted speech rely on processing artifacts specific to a particular synthesis or voice conversion algorithm \cite{ASVsurvey} such as phase information. Phase domain features outperform magnitude related features, because spoofed speech doesn't retain the natural phase information.\\
The most commonly used phase-based features are related to group delay information. First of them are \textbf{\emph{Group Delay}} (GD) features. Group delay is defined as derivative of the phase spectrum along the frequency axis \cite{D}.
The described way of calculation of the group delay function at frequency bins near zeros, that can occur near the unit circle, will results in high amplitude false peaks. These peaks mask out the formant structure. Due to this fact, \textbf{\emph{Modified Group Delay}} (MGD) function suppress these zeros by the use of cepstrally smoothed magnitude spectrum instead of the original version. MGD features are known as more stable in speach recognition and were mostly used by participants of the Challenge. They were implemented in spoofing detection systems in \cite{D}, \cite{E}, \cite{F}, \cite{G} and \cite{I}. However experiments from \cite{D} presented in Table \ref{tab:MGDF}, demonstrate that MGD are not so effective as GD are for spoofing detection task.
\begin{table}[]
    \centering
    \begin{tabular}{|c|c|c|c|c|c|c|}
    \hline
    & \multicolumn{6}{ c| }{Spoofing type}\\
    \cline{2-7}
    Features type& S1 & S2 & S3 & S4 & S5 & All\\
    \hline	
    MS&0.347&0.254&0.054&0.054&1.603&0.543\\
    RLMS&0.000&0.093&0.039&0.039&1.456&0.486\\
    \hline
    GD&0.054&0.054&0.039&0.000&0.161&0.114\\
    MGD&1.148&2.311&0.147&0.147&2.311&1.572\\
    \hline
    IF&0.161&0.401&0.147&0.147&0.948&0.428\\
    BPD&2.243&4.955&0.401&0.347&5.155&3.431\\
    \hline
    \end{tabular}
    \caption{Experiments results for the MLP-system with different features for different spoofing types obtained on the development dataset (EER \%)}
    \label{tab:MGDF}
\end{table}
Another approach to solve the problem of GD were used by \cite{E}. They implemented the \textbf{\emph{Product Spectrum}} based features that were calculated as the product of power spectrum and GD function, thus combining information from amplitude and phase spectra (PC-MFCC).\\
The second feature type, mitigating the effect of zeros in group delay, was \textbf{\emph{All-pole Group Delay-based}} features (WLP-GDCC). The main idea of this method is to keep only the vocal tract component of the speech signal and discard the contribution of the excitation source.\\
\begin{table}[]
    \centering
    \begin{tabular}{|c|c|c|c|}
    \hline
    & \multicolumn{3}{ c| }{Spoofing type}\\
    \cline{2-4}
    Features type& Known attacks & Unknown attacks & All\\
    \hline
    MGD&1.924&7.124&4.524\\
    PS-MFCC&0.652&5.372&3.011\\
    WLP-GDCC&1.436&8.941&5.188\\
    \hline
    \end{tabular}
    \caption{Experiments results for the GMM-system with different features for different spoofing types obtained on the development dataset (EER \%)}
    \label{tab:MGDCC}
\end{table}
As the phase changes depending on the splitting position of the input utterance it is important to normalize obtained phase information. \cite{I} and \cite{D} use \textbf{\emph{Relative Phase}} extraction methods to reduce phase variation. These approaches have differences but both are based on the pitch synchronization of the slitting section instead of using fixed frame. Authors of \cite{I} obtained impressive results. Comparing this features with MGD features on the base of one system researched obtained 0.83\% EER for MGD features and 0.013\% EER for Relative Phase features on the development set \cite{I}, while the perfomance of system based on Pitch Synchronous Phase(PSP) features \cite{D} is slightly less than for MGD-based one.
These feature type was also used in papers \cite{J} and \cite{K}.\\
Another method to extract phase information is to use \textbf{\emph{Instantaneous frequency}} (IF) estimation. While group delay is the derivative of the phase along the frequency axis, instantaneous frequency can be calculated as the derivative of the phase along the time axis. IF features are used in spoofing detection systems in \cite{A} and \cite{D}.\\
Researchers in \cite{D} also used \textbf{\emph{Baseband Phase Difference}} (BPD) from \cite{BPD} as more stable time-derivative phase based features and found out that these features contain different artifacts from the IF features. However, their results, presented below in Table \ref{tab:MGDF}, demonstrate that BDF features are not so efficient as IF features are, especially for voice conversion techniques S2 and S5.\\
In our system \cite{B} we used \textbf{\emph{CosPhasePC}} features which were extracted from unwrapped phase spectrum by applying cosine normalization and dimensionality reduction by means of principal components analysis. Results for system using CosPhasePC features on the development data set, presented in Table \ref{tab:MWPC}, confirms that CosPhasePc features are highly effective for all known types of spoofing attacks. Similar features was also used by \cite{E} and \cite{F}. Experiments in \cite{E} confirm the power of cosine normalized phase-based features for known attacks.
\subsubsection{Local Binary Patterns.}
Authors of \cite{F} investigated the possibility to use spectra-temporal structure for spoofing detection task. In order to do this they used Local Binary Patterns (LBP) approach proposed for texture recognition. The spectrogram was used as acoustic representation. Authors treated it as 2D image to apply uniform LBP features extraction. Authors noticed that despite the traditional LBP algorithms for images, here they derived the histogram over each coefficient separately and used unique LPS without rotation invariance. Thus, they used the texture of the spectral magnitude as features to detect spoofed speech. By using these features they achieved 0.858 \% EER for all spoofing attacks from the development set.
\subsubsection{Wavelet transform.}
In our work, in order to include detailed time-frequency analysis of the speech signal in spoofing detection countermeasures, we proposed features based on applying the multiresolution wavelet transform \cite{Wavelet}, that was adapted to the mel scale, called \textbf{\emph{Mel Wavelet Packet Coefficients}}. We used Daubechies wavelets db4 in the wavelet-decomposition. Using Teager Keiser Energy Operator instead of classical energy of the frequency sub-band makes these features more informative and noise-robust than classical sample energy. We also applied projection on the eigenvector basis for features decorrelation. Our experiments on the development set (Table \ref{tab:MWPC}) demonstrated that described features showed the best results in terms of performance of individual system based on concrete feature type.\\
\begin{table}[]
    \centering
    \begin{tabular}{|c|c|c|c|c|c|c|}
    \hline
    & \multicolumn{6}{ c| }{Spoofing type}\\
    \cline{2-7}
    Features type& S1 & S2 & S3 & S4 & S5 & All\\
    \hline
    MFCC&0.38&2.13&0.36&0.39&1.48&1.14\\
    MFPC&0.13&0.29&0.09&0.09&0.37&0.23\\
    \hline	
    CosPhasePC&0.13&0.20&0.04&0.05&0.23&0.15\\
    MWPC&0.03&0.11&0.00&0.00&0.08&0.05\\
    \hline
    \end{tabular}
    \caption{{Experiments results for the TV-SVM system with different features obtained on the development dataset (EER \%)}}
    \label{tab:MWPC}
\end{table}
Authors of \cite{A} proposed auditory-based cepstral coefficients called \textbf{\emph{Cochlear Filter Cepstral Coefficients}} (CFCC). They can be extracted by applying the cochlear filterbank based on auditory trasform, hair cell function, nonlinearity and discrete cosine transform. A brief description of the feature extraction procedure is presented in \cite{A}. Authors use CFCC features together with IF features, described above, to combine both envelope structure and IF information (CFCCIF). Framewise IF features are multiplied with the corresponding nerve spike density envelope, obtained during the CFCC extraction operation. Thus, IF obtained in silence regions will be suppressed. The derivative operation is used to capture the changing information in envelope and IF for consecutive frames. In \cite{A} researchers obtained 2.6 \% EER for CFCC-based and 1.4 \% EER for CFCCIF-based individual systems. Comparison with 2.66 \% EER for MFCC-based system shows that CFCCIF features are highly effective for spoofing detection task.

\subsubsection{Phonetic level.}
\label{sec:PPP}
Based on the achievements of \cite{PPP1} authors of \cite{G} proposed to use combination of MFCC with the \textbf{\emph{Phonetic Level Phoneme Posterior Probability}} (PPP) tandem features for spoofing detection task. They used multilayer perceptron based phoneme recognizer with a English acoustic model trained on the TIMIT database for phoneme decoding and obtained 1.72 \% EER on their SVM based spoofing detection system. That is expressive improvement in comparison with 8.46 \% EER for MFCC features on the similar system.

\section{High level features extraction}
In our work for the acoustic space modelling we used the standard Total Variability approach, which is widely used in speaker verification systems \cite{TV2}. The main idea of this approach consists in finding a low dimensional subspace of the GMM supervector space, named the total variability space that represents both speaker and channel variability. The vectors in the low-dimensional space called super-vectors or i-vectors. These i-vectors were extracted by means of Gaussian factor analyser defined on mean supervectors of the Universal Background Model (UBM) and Total Variability matrix T. UBM was represented by the diagonal covariance Gaussian mixture models of the used features. This approach was also used by \cite{F}, \cite{G}, while \cite{A} used two simple GMM models for natural and spoofed speech.\\
Systems from \cite{C} used Deep Neural Network (DDN) models. The mean values of outputs of last hidden layer from the trained  neural network were used as a final representation of the signal s, which are new robust representations, called spoofing vectors (s-vector).

\section{Back-end}
\subsubsection{GMM.} 
Most part of the participants used standard GMM-classifiers in their systems. These are \cite{A}, \cite{E}, \cite{F}, \cite{I}, \cite{K}.
\subsubsection{SVM.} Support Vectors Machine (SVM) was the second popular classifier in the ASVspoof challenge. We used SVM with linear kernel in our primary system as it presented the best performance in our experiments. To train SVN we used the efficient LIBLINEAR \cite{liblinear} library with default C-values equal to 1. Authors of \cite{F}, \cite{G} and \cite{J} also chose SVM as classifiers in their submitted systems.
\subsubsection{DNN.}
System performed in \cite{D} combine all 6 proposed features types. Moreover the feature vectors were concatenating within a window to incorporate long term temporal information. In  order to handle  the high demensional feature vectors authors used Deep Neural Network with one hidden layer. This system achieved 0.001 \% EER on all spoofing types from the development set.\\
Authors in \cite{C} investigated Deep Neural Network classifier for two types of constructions: 6 classes(individual for each type of spoofing attack) and 2 classes (1 class for all spoofing attacks). They obtained better results for DNN than for GMM classifier. And although 2 classes DNN classification performed better that 6 classes configuration on the development set, their small scaled experiments convinced that 6 classes classification has better performance on the unknown spoofing attacks.
\subsubsection{DBN.} In our system we used classifier based on Deep Belief Network with softmax output units and stochastic binary hidden units.  We used layer-wise pretreating of the layers by means of Restricted Boltzmann Machines (RBMs) and then applied back-propagation to train the DBN in a supervised way to perform classification. However our experiments on the development set demonstrated that linear SVM classifier works better on the proposed features. In this system we probably failed to avoid the effects of the stronger overfitting on these training dataset, in comparison with SVM.
\subsubsection{K-nearest.}
Authors of \cite{G} compared several classification approaches: K-nearest neighbor classification (KNN) with 2 classes for human and spoofed speech, cosine similarity scoring, simplified PLDA classifier with 6 classes (individual class for each spoofing type), two stage LDA with 2 subspace (speaker subspace and spoofing subspace) and SVM as 2 class classification. 
\begin{table}[]
    \centering
    \begin{tabular}{|p{1.8cm}|p{1.8cm}|p{1.8cm}|p{1.8cm}|p{1.8cm}|p{1.8cm}|}
    \hline
    Linear kernel SVM& Polynomial kernel SVM& Cosine scoring & KNN & Simplified PLDA & Two stage PLDA\\
    \hline	
    1.86 & 1.06 & 2.86 & 2.46 & 1.89 & 10.18\\
    \hline
    \end{tabular}
    \caption{EER for spoofing detection systems based on different classifiers for system from \cite{G} (\%)}
    \label{tab:classifiers}
\end{table}
Table \ref{tab:classifiers} demonstrates results obtained for their system based on MFCC and PPP features described in \ref{sec:PPP} with score-fusion. According to these results SVM classifiers outperform the others.
\section{Evaluation results}
A comparison of all final systems of the participants is possible only on the evaluation data set. These experiments results were presented in \cite{ASV} and described below in Table \ref{tab:ASV}. The best results in terms of unknown attacks and average was obtained by system based on score-level fusion of MFCC and CFCCIF features, GMM modeling and log-likelihood scoring. Many systems used total variability modelling for high level features extraction, which also improve the performance of spoofing detection. Talking about classifiers, it should be mentioned that it is highly complicated to define the best classifiers based on the evaluation results of the ASVspoof Challenge because it depends also on the pre-processing effect, type of features, modelling type and on the details of the classification task (2 class classification or 6 class classification with each class for each spoofing attack type). Nevertheless we can figure out the strong success of SVM classifiers, that was confirmed by several researches, and high performance on neural networks for spoofing detection task. Probably, further study in this field will lead to more significant results.
\begin{table}[]
    \centering
    \begin{tabular}{|c|c|c|c|}
    \hline
    & \multicolumn{3}{ c| }{Equal Error Rates (EERs)}\\
    \cline{2-4}
System ID & Known attacks & Unknown attacks & Average\\
\hline
A \cite{A} & 0.408 & \textbf{2.013} & 1.211\\
B \cite{B} & 0.008 & 3.922 & 1.965\\
C \cite{C} & 0.058 & 4.998 & 2.528\\
D \cite{D} & \textbf{0.003} & 5.231 & 2.617\\
E \cite{E} & 0.041 & 5.347 & 2.694\\
F \cite{F} & 0.358 & 6.078 & 3.218\\
G \cite{G} & 0.405 & 6.247 & 3.326\\
H & 0.670 & 6.041 & 3.355\\
I \cite{I} & 0.005 & 7.447 & 3.726\\
J \cite{J} & 0.025 & 8.168 & 4.097\\
K \cite{K} & 0.210 & 8.883 & 4.547\\
    \hline
    \end{tabular}
    \caption{Evaluation results of ASVspoof Challenge 2015 (EER \%)}
    \label{tab:ASV}
\end{table}
Anti-spoofing system, presented in \cite{D}, that used 6 different types of features, including magnitude based, GD and MDG, IF and PSP features and used MLP classification was the best system for known types of attack, while it achieved only 4th result in terms of unknown types of attacks.
Our primary system, based on the MFCC, MFPC and CosPhasePC feature-level fusion, TV modelling and SVM classifier achieved the second place with a stable 2nd results for known and unknown spoofing attacks.\\
All proposed systems perform poor performance for S10 type of spoofing attack. This fact leaves the problem of efficient spoofing detection countermeasures to be actual for further investigations.

\section{Conclusion}
In this paper we investigated modern tendencies in spoofing detection on the base of ASVspoof Challenge 2015 results. Experimental results of the participants, confirm that the most efficient systems use several types of features, responsible for different information and artifacts of the speech signal. Because these systems can catch complementary information that is not evident for individual feature-based systems. Most often these features contain magnitude and phase information. However, phoneme features also were highly effective.\\
Several preprocessing techniques were found out to be crucial for concrete features type. For example, MFCC features are sensitive to pre-emphasis step, and it can be helpful with fine tuned parameters. According to our experiments VAD may throw out informative artifacts locating between speech fragments.\\
Classification comparisson show that SVM is highly efficient for spoofing detection task, as well as neural network approaches.

\textbf{Acknowledgements.} This work was financially supported by the Ministry of Education and Science of the Russian Federation, Contract 14.578.21.0126 (ID RFMEFI57815X0126).


\begin{thebibliography}{}
\bibitem{villalba} Villalba E., Lleida E.: Speaker verification performance degradation against spoofing and tampering attacks, in Proc. of the FALA 2010 Workshop, pp. 131–134, 2010.


\bibitem{ASV} Zhizheng Wu, Tomi Kinnunen, Nicholas Evans, Junichi Yamagishi, Cemal Hanilc, Md Sahidullah, Aleksandr Sizov: ASVspoof 2015: the First Automatic Speaker Verification Spoofing and Countermeasures Challenge 2015. Available online: \url{http://www.spoofingchallenge.org/is2015_asvspoof.pdf}




\bibitem{A} Tanvina B. Patel, Hemant A. Patil: Combining Evidences from Mel Cepstral, Cochlear Filter Cepstral and Instantaneous Frequency Features for Detection of Natural vs. Spoofed Speech, Interspeech 2015.
\bibitem{B} Sergey Novoselov, Alexandr Kozlov, Galina Lavrentyeva, Konstantin Simonchik, Vadim Shchemelinin: STC Anti-spoofing Systems for the ASVspoof 2015 Challenge, arXiv:1507.08074, 2015.
\bibitem{C} Nanxin Chen, Yanmin Qian, Heinrich Dinkel, Bo Chen, Kai Yu: Robust Deep Feature for Spoofing Detection - The SJTU System for ASVspoof 2015 Challenge, Interspeech 2015

\bibitem{D} Xiong Xiao, Xiaohai Tian, Steven Du, Haihua Xu, Eng Siong Chng, Haizhou Li:Spoofing Speech Detection Using High Dimensional Magnitude and Phase Features: the NTU Approach for ASVspoof 2015 Challenge, Interspeech 2015

\bibitem{E} Md Jahangir Alam, Patrick Kenny, Gautam Bhattacharya, Themos Stafylakis: Development of CRIM System for the Automatic Speaker Verification Spoofing and Countermeasures Challenge 2015, Interspeech 2015

\bibitem{F} Yi Liu, Yao Tian, Liang He, Jia Liu, Michael T. Johnson: Simultaneous Utilization of Spectral Magnitude and Phase Information to Extract Supervectors for Speaker Verification Anti-spoofing, Interspeech 2015

\bibitem{G} Shitao Weng, Shushan Chen, Lei Yu, Xuewei Wu, Weicheng Cai, Zhi Liu, Ming Li: The SYSU System for the Interspeech 2015 Automatic Speaker Verification Spoofing and Countermeasures Challenge, arXiv:1507.06711, 2015

\bibitem{I} Longbiao Wang , Yohei Yoshida, Yuta Kawakami, Seiichi Nakagawa: Relative phase information for detecting human speech and spoofed speech, Interspeech 2015

\bibitem{J} Jesus Villalba, Antonio Miguel, Alfonso Ortega, Eduardo Lleida: Spoofing Detection with DNN and One-class SVM for the ASVspoof 2015 Challenge, Interspeech 2015

\bibitem{K} Jon Sanchez, Ibon Saratxaga, Inma Hernaez, Eva Navas, Daniel Erro: The AHOLAB RPS SSD Spoofing Challenge 2015 submission, Interspeech 2015.

\bibitem{liblinear} LIBLINEAR: A library for Large Linear Classification. \url{https://www.csie.ntu.edu.tw/~cjlin/liblinear/}

\bibitem{VAD_E1} T.Kinnunen, P.Rajan: A practical, self adaptive voice activity detector for speaker verification with noisy telephone and microphone data, in Proc. of ICASSP, pp. 7229-7233, 2013.

\bibitem{handbook} Sebastien Marcel, Mark S. Nixon,Stan Z. Li: Handbook of Biometric Anti-spoofing: Trusted Biometrics Under Spoofing Attacks. In: Springer (2014)




\bibitem{ASVsurvey} Z. Wu , N. Evans, T. Kinnunen, J. Yamagishid, F. Alegreb, H. Lia: Spoofing and countermeasures for speaker verification: a survey, Speech Communication, vol. 66, February 2015, pp. 130–153.

\bibitem{BPD} M.Krawczyk, T.Gerkmann: Shift phase reconstruction in voiced speech for an improved single-channel speech enhancement, IEEE/ACM Transactions on Audio, Speech and Language Processing(TASLP), vol.22, no. 12, pp. 1931-1940, 2014

\bibitem{Wavelet} S. Mallat: A wavelett Tour of Signal Processing, 3rd ed, Academic Press, 2008

\bibitem{PPP1} L.D'Haro, R.Cordoba, C.Salamea, J.Echeverry: Extended phone log-likelihood ratio features and acoustic-based i-vectors for languages recognition, in Proc. ICASSP. IEEE, 2014, pp. 5379-5383.



\bibitem{TV2} S. Novoselov, T. Pekhovsky, K. Simonchik: STC Speaker Recognition System for the NIST i-Vector Challenge in Proc. Odyssey 2014 - The Speaker and Language Recognition Workshop, 2014.


\bibitem{DBN} Hinton, G. E., Osindero, S., Teh, Y: A fast learning algorithm for deep belief nets, Neural Computation, Vol. 18, pp. 1527–1554, Jul. 2006
\end{thebibliography}
\end{document}